\documentclass[conference]{IEEEtran}
\usepackage{fancyhdr}
\IEEEoverridecommandlockouts
\usepackage{cite}
\usepackage{amsmath,amssymb,amsfonts}
\usepackage{graphicx}
\usepackage{textcomp}
\usepackage{xcolor}
\usepackage[hidelinks]{hyperref} 
\usepackage{cleveref}

\usepackage{multirow}
\usepackage{inconsolata}
\usepackage{algorithm}
\usepackage{algpseudocode}
\usepackage{subfig}
\usepackage{threeparttable}
\floatname{algorithm}{Code}
\usepackage[english]{babel}
\usepackage{soul}
\def\BibTeX{{\rm B\kern-.05em{\sc i\kern-.025em b}\kern-.08em
    T\kern-.1667em\lower.7ex\hbox{E}\kern-.125emX}}
\begin{document}

\newcommand{\copytmp}[1]{\textcolor{red}{#1}}
\newcommand{\notsure}[1]{\textcolor{blue}{#1}}
\newcommand{\todo}[1]{\textcolor{purple}{#1}}
\newcommand{\comment}[1]{\textcolor{brown}{(#1)}}
\newcommand{\controlline}[3]{%
  \ifthenelse{\equal{#1}{1}}{#2}{#3}%
}

\title{UNR: Unified Notifiable RMA Library for HPC}


    
    
    
    
    
    
    
    

\author{
    \IEEEauthorblockN{
        Guangnan Feng\IEEEauthorrefmark{2},
        Jiabin Xie\IEEEauthorrefmark{2},
        Dezun Dong\IEEEauthorrefmark{3},
        Yutong Lu\IEEEauthorrefmark{2}
    }
    \IEEEauthorblockA{
        \IEEEauthorrefmark{2}Sun Yat-sen University, Guangzhou, China
    }
    \IEEEauthorblockA{
        \IEEEauthorrefmark{3}National Laboratory for Parallel and Distributed Processing (PDL), Changsha, China
    }
    \IEEEauthorblockA{fenggn3,xiejb6@mail2.sysu.edu.cn, 13787783335@139.com, luyutong@mail.sysu.edu.cn}
}

\maketitle

\thispagestyle{fancy}
\lhead{}
\rhead{}
\chead{}
\lfoot{\footnotesize{
SC24, November 17-22, 2024, Atlanta, Georgia, USA
\newline 979-8-3503-5291-7/24/\$31.00 \copyright 2024 IEEE}}
\rfoot{}
\cfoot{}
\renewcommand{\headrulewidth}{0pt}
\renewcommand{\footrulewidth}{0pt}

\begin{abstract}
Remote Memory Access (RMA) enables direct access to remote memory to achieve high performance for HPC applications.
However, most modern parallel programming models lack schemes for the remote process to detect the completion of RMA operations. 
Many previous works have proposed programming models and extensions to notify the communication peer, but they did not solve the multi-NIC aggregation, portability, hardware-software co-design, and usability problems. 
In this work, we proposed a Unified Notifiable RMA (UNR) library for HPC to address these challenges.
In addition, we demonstrate the best practice of utilizing UNR  within a real-world scientific application, PowerLLEL.
We deployed UNR across four HPC systems, each with a different interconnect.
The results show that PowerLLEL powered by UNR achieves up to a 36\% acceleration on 1728 nodes of the Tianhe-Xingyi supercomputing system.

\end{abstract}

\begin{IEEEkeywords}
Remote Memory Access, Communication Optimization, Synchronization, Parallel Programming Model, \controlline{1}{MPI}{Message Passing Interface}
\end{IEEEkeywords}

\section{Introduction}

Remote Memory Access (RMA) is an important technique for high-performance parallel programming.
RMA allows a process to directly access the memory of a remote process without the involvement of the remote processor and the operating system, which can significantly reduce the access latency and improve the communication performance
\cite{krishnanScalingLinearAlgebra2010, straatsmaEliminatingSynchronousCommunication2013, xieExtremescaleDirectNumerical2024, dingLeveragingOneSidedCommunication2020, sunOptimizingFinegrainedCommunication2012}.

However, most modern parallel programming models do not provide schemes for a remote process to detect the completion of each RMA operation\cite{MPIForum,bonacheaUPCLanguageSpecifications2013,reidCoarraysNextFortran2010,nieplochaARMCIPortableRemote1999,bauerLegionExpressingLocality2012,WelcomeHPXDocumentation,leeImplementationPerformanceEvaluation2010}.
Dedicated synchronization interfaces are designed to notify the remote process that all the previous RMA operations have finished.
Such designs impede computation-communication pipeline optimization: the receiver cannot immediately consume a portion of the received data but must synchronize with the sender.
Many RMA-enabled Network Interface Cards (NICs) and their low-level network programming interfaces provide schemes for remote processes to poll the completion of RMA operations\cite{RdmacoreLibibverbsMaster,fujitsuDevelopmentStudioUTofu2021,enterpriseGNIAPIReference,kumarPAMIParallelActive2012,hudsonPortalsMessagePassing2006}. 
Multiple programming models and extensions to existing models have been proposed to provide the ability to notify a remote process that an RMA operation is finished\cite{OpenSHMEMSpecification2020,breitbartEvaluationGlobalAddress2014,belliNotifiedAccessExtending2015,sergentEfficientNotificationsMPI2019,bonacheaEfficientPointtopointSynchronization2006,dinanReducingSynchronizationOverhead2014,fanfarilloNotifiedAccessCoarray2017,NotifiedAccessCoarraybased2018,kisselPhotonRemoteMemory2016,gysiDCUDAHardwareSupported2016,liuMPINewCommunication2021,wickramasingheRDMAManagedBuffers2019}. 
Nevertheless, none of these works addresses the following challenges simultaneously:

\begin{itemize}

\item \textbf{Multi-NIC aggregation}:
Each computing node with multiple NICs is increasingly prevalent in HPC systems.
Multi-NIC aggregation can reduce flow completion time by transmitting segmented sub-messages through multiple NICs.
The multi-NIC aggregation in RMA requires an efficient and portable mechanism to inform the completion of all the sub-messages with limited resources.

\item \textbf{Portability}:
Since there are different low-level network programming interfaces,
a unified interface requires comprehensive interface analysis, classification, and implementation for different platforms to achieve both code portability and performance portability.

\item \textbf{Software-hardware Co-design}: 
To prevent the NIC's event queue from overflowing, a dedicated polling thread is necessary to continuously monitor the NIC. However, this thread may compete for computational resources with other computing tasks.
We urgently need a hardware-software co-design to replace the polling thread.

\item \textbf{Usability}:
When switching from classical two-sided communication operations to one-sided RMA operations, 
synchronization errors and remote offset calculation errors are two common issues.
There is a strong need for solutions to help users avoid these bugs.
\end{itemize}

To address the above challenges, we proposed a \textbf{U}nified \textbf{N}otifiable \textbf{R}MA (\textbf{UNR}) library for HPC, which contains the following features:

\begin{itemize}

\item \textbf{M}ulti-channel\,\textbf{M}ulti-message\,\textbf{A}ggregated\,\textbf{S}ignal\,(\textbf{MMAS}) offers multi-NIC aggregation and aggregation of multiple notifications from one or more peers. It efficiently utilizes the custom bits in low-level network interfaces to transmit signals.

\item \textbf{UNR support levels} achieve portability and hardware-software co-design. We classify the ability of a NIC to support UNR into multiple levels and define implementation specifications for each level. The highest support level does not have hardware support yet, but it guides the hardware design of future high-performance NICs.

\item \textbf{Bug-avoiding interfaces} provide usability by automatically detecting synchronization errors when users call the corresponding interfaces, and replacing remote address calculation with transportable data handle.

\end{itemize}

In addition, we demonstrate a best practice of utilizing UNR to optimize a real-world scientific application, PowerLLEL. In the hotspots of the application, we remove all the pre-synchronization by communication dependency analysis and all the post-synchronization by UNR.

We evaluated UNR using micro-benchmarks and PowerLLEL across four HPC systems to show the performance and portability.
The results demonstrate that PowerLLEL exhibits performance enhancement on all four platforms, achieving up to a 36\% speedup on 1728 nodes of the Tianhe-Xingyi supercomputing system.
The rest of this paper is organized as follows:
\Cref{sec:background} introduces RMA operations.
\Cref{sec:related_works} discusses previous works.
In \Cref{sec:design}, we propose the UNR library design.
In \Cref{sec:best_practice}, we show a best practice of UNR\controlline{1}{ with a real scientific application, PowerLLEL}.
In \Cref{sec:evaluation}, we evaluate UNR on four different HPC systems.
Finally, \Cref{sec:conclusion} and \ref{sec:future_work} present our conclusions and future work.

\section{Background} \label{sec:background}

\begin{figure}[t!]
    \centering
    \subfloat[Eager Protocol\label{fig:eager}]{\includegraphics[width=0.5\linewidth]{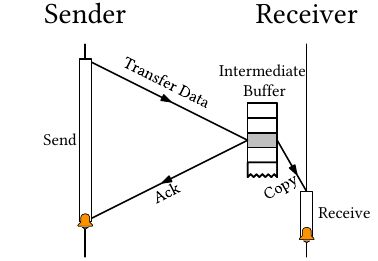}}
    \subfloat[Rendezvous Protocol\label{fig:rend}]{\includegraphics[width=0.5\linewidth]{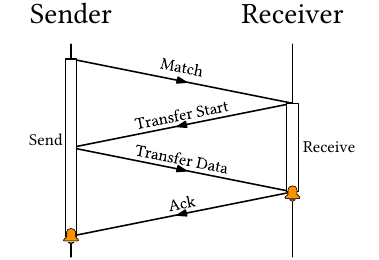}}
    \hfill
    \subfloat[PUT Operation\label{fig:PUT}]{\includegraphics[width=0.5\linewidth]{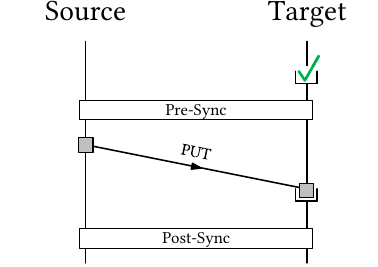}}
    \subfloat[GET Operation\label{fig:GET}]{\includegraphics[width=0.5\linewidth]{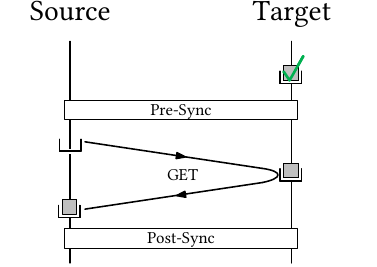}}
    \hfill
    \subfloat[Notified PUT Operation\label{fig:NPUT}]{\includegraphics[width=0.5\linewidth]{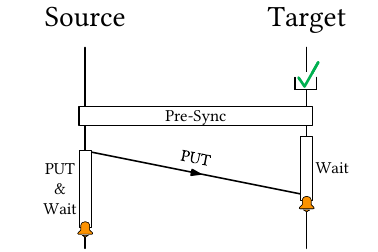}}
    \subfloat[Notified GET Operation\label{fig:NGET}]{\includegraphics[width=0.5\linewidth]{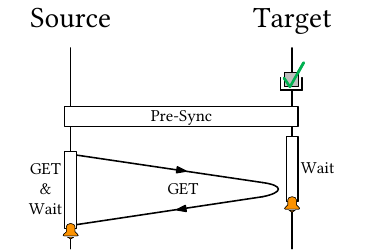}}
    \hfill
    \includegraphics[width=1.0\linewidth]{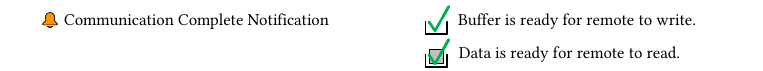}
    \caption{Communication Protocols and Operations}
    \label{fig:eager_rend}
\end{figure}

Point-to-point communications are categorized into one-sided and two-sided communications.
In two-sided communication, both the sender and the receiver are involved in communication.
The sender has to specify the address of the sending buffer, and the receiver has to set the address for receiving data.
Two-sided communication can be implemented by Eager or Rendezvous Protocol, as shown in \Cref{fig:eager} and \ref{fig:rend}, respectively. Both protocols require additional data copy or handshake.

In one-sided communications (RMA operations), the sender can use PUT (RMA Write) to write the data directly to the receiver's memory without the involvement of the receiver, as shown in \Cref{fig:PUT}. The receiver can use GET (RMA read) to fetch the data directly from the sender's memory without the involvement of the sender, as shown in \Cref{fig:GET}. 
In this paper, we use uppercase PUT and GET to represent these one-sided operations.

As a price of removing additional data copy or handshake\controlline{1}{}{ in each message},
one-sided operations need to be used with pre-synchronization to ensure the buffer or data is ready on the remote, and post-synchronization to confirm the completion of operations.
It is unsafe to take the arrival of partial bytes as an indication of receiving the complete message, as fragments of a message may arrive out of order due to adaptive routing\cite{glassTurnModelAdaptive1992} and multi-rail network\cite{wangMRtreeParametricFamily2021,wolfePreliminaryPerformanceAnalysis2017}, i.e. a network where fragments of a message may transmitted by multiple NICs in a node. 

Using additional synchronizations is an easy and portable way to confirm the completion of operations.
An explicit synchronization is performed after one or more RMA operation(s) to inform the target process that the operation(s) by the remote process(es) are finished.
Most parallel programming models use these schemes and explicitly differentiate data movement and synchronization, including 
MPI\cite{MPIForum}, 
UPC\cite{bonacheaUPCLanguageSpecifications2013},
Co-Array Fortran\cite{reidCoarraysNextFortran2010}, 
ARMCI\cite{nieplochaARMCIPortableRemote1999},
Legion\cite{bauerLegionExpressingLocality2012}, 
HPX\cite{WelcomeHPXDocumentation}, and
XcalableMP\cite{leeImplementationPerformanceEvaluation2010}.
It is the same for communication middlewares, including
UCX\cite{shamisUCXOpenSource2015}, 
GASNET\cite{bonacheaGASNetEXHighPerformancePortable2019}, and
CH3/4\cite{mpichHighPerformancePortableMPI2022}.
If a synchronization is used to confirm the completion of many RMA operations, its cost is usually negligible.

However, if the remote process needs to know the completion of each RMA operation, e.g., in a producer-consumer pattern, the overhead of an additional synchronization for each operation is high\cite{belliNotifiedAccessExtending2015}.
A direct way to solve this problem is using low-level network programming interfaces, which can produce complete events to inform both the sender and the receiver a one-sided operation is complete, as shown in \Cref{fig:eager_rend}e-f.
We call it Notifiable RMA Primitives in this work.

To the best of our knowledge, all modern high-performance interconnect systems support Notifiable RMA Primitives, as shown in \Cref{table:NIC_Level}.
The Notifiable RMA Primitives of these interconnect systems offer users custom bits of varying lengths, which can be utilized to differentiate the completion of RMA operations.
This work unifies these Notifiable RMA Primitives on different platforms to provide a portable and powerful assistance communication library. 

\section{Related Works} \label{sec:related_works}

Since using Notifiable RMA Primitives in applications is difficult for users, it is a common idea to provide a scheme with a combination of RMA operation and notification signal.
Many previous works have shown significant performance improvement in benchmarks, proxy applications, and real applications after adopting Notifiable RMA Interfaces. 

OpenSHMEM\cite{OpenSHMEMSpecification2020}, GASPI\cite{breitbartEvaluationGlobalAddress2014} have added Notifiable RMA Interfaces into their specifications.
However, they do not guarantee that Notifiable RMA Interfaces are implemented by Notifiable RMA Primitives,
which means Notifiable RMA Interfaces may be implemented by a one-sided communication operation followed by a synchronization operation.

For usability, 
foMPI (a non-standard MPI implementation for CRAY)\cite{belliNotifiedAccessExtending2015}, 
MPIX (extension for OpenMPI)\cite{sergentEfficientNotificationsMPI2019},
Extended UPC (non-standard)\cite{bonacheaEfficientPointtopointSynchronization2006},
Extended SHMEM (non-standard)\cite{dinanReducingSynchronizationOverhead2014},
and notified access enabled Co-Array Fortran\cite{fanfarilloNotifiedAccessCoarray2017, NotifiedAccessCoarraybased2018}
are proposed to extend standard interfaces for existing programming models.
Photon\cite{kisselPhotonRemoteMemory2016}, dCUDA\cite{gysiDCUDAHardwareSupported2016}, RDMO\cite{liuMPINewCommunication2021} are proposed as new programming models or middleware.
RDMA Managed Buffers\cite{wickramasingheRDMAManagedBuffers2019} is an RDMA transport layer that co-exists with other programming models, such as MPI.

There are three schemes to notify the remote process in previous works' Notifiable RMA Interfaces.
In \cite{breitbartEvaluationGlobalAddress2014, grunewaldGASPIAPISpecification2013, wickramasingheRDMAManagedBuffers2019, wickramasingheEnablingEfficientInterNode2018, sergentEfficientNotificationsMPI2019}, receiver has to collect each notification after its arrival, which may cause unnecessary memory access.
\cite{bonacheaEfficientPointtopointSynchronization2006} attaches the notification to each memory region, but the number of \controlline{0}{available }{}registered memory regions is limited in some systems.
\cite{belliNotifiedAccessExtending2015, gysiDCUDAHardwareSupported2016} propose counted notifications that only notify users after $n$ matching accesses are performed. They divide the 32 custom bits into a 16-bit source rank and a 16-bit tag, resulting in only $65536$ available ranks and $65536$ tags.

Considering adding Notifiable RMA Interfaces into MPI-5.0 is one of the 148 items on its to-do list\cite{RMANotificationIssue2024}, but the issue is still under discussion due to the processing needed for the event queue.
\cite{belliNotifiedAccessExtending2015} and \cite{gysiDCUDAHardwareSupported2016} suggest integrating the notification infrastructure within hardware, but do not discuss the required hardware features.
RVMA\cite{grantRVMARemoteVirtual2021} proposed a \controlline{0}{new}{} hardware-software co-design, which includes a \controlline{0}{lightweight}{}completion notification mechanism. However, a redesign of NIC hardware is required. 

\section{UNR Design} \label{sec:design}

\subsection{Overview}

UNR is a one-sided communication acceleration library for modern HPC systems.
This library aims to unify the different Notifiable RMA Primitives on different HPC interconnects for applications' portability, performance, and usability.

The UNR library consists of two main components: UNR Transport Layer and UNR Interface Module.
The UNR Transport Layer abstracts the differences across various Notifiable RMA Primitives.
It currently supports GLEX Channel, Verbs Channel, and a fallback MPI Channel.
Although we have not implemented other channels due to limited access to these systems, we provide a comprehensive analysis in \Cref{sec:UNR_level} to \controlline{0}{demonstrate}{show} that the UNR Transport Layer can support almost all modern HPC interconnects.
The UNR Interface Module offers Notifiable RMA Interfaces for users and schedules transmission requests across multiple UNR Transport Channels.

The key design of UNR is Multi-channel Multi-message Aggregated Signal, which contributes to the performance, usability, and portability of UNR.
Regarding performance, it can utilize the bandwidth from multiple NICs.
On the usability front, users can verify the receipt of multiple messages from one or multiple sources with a single signal, 
i.e., the multi-channel and multi-message aggregation is transparent to users.
For portability, MMAS determines how to abstract different Notifiable RMA Primitives across various platforms. The UNR support levels are defined according to how well a NIC can support the aggregation.

For \controlline{0}{better}{}usability, UNR can co-exist with other parallel programming models and provide bug-avoiding interfaces. 
Users can gradually replace the original communication operations in the performance bottleneck with the operations provided by UNR, which reduces the difficulty of adopting UNR and testing the new implementation. 
UNR provides interfaces that avoid synchronization errors and the calculation of the remote address offsets, which often cause bugs when using RMA.




\subsection{MMAS: Multi-channel Multi-message Aggregated Signal} 
\label{sec:MMAS}

\begin{figure}[b]
  \centering
  \includegraphics[width=0.982\linewidth]{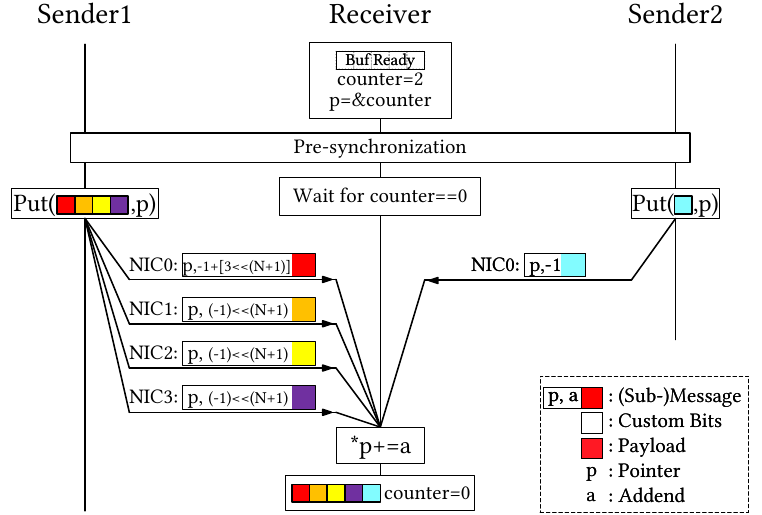}
  \caption{\textbf{Multi-channel Multi-message Aggregated Signal}. The receiver waits for messages from two senders until the \texttt{counter} reaches $0$.
Sender1 divides the large message into four sub-messages and transfers them through four NICs.}
  \label{fig:MNMMAS}
\end{figure}

\begin{table*}[t]
\begin{threeparttable}[b]
\centering
\renewcommand{\arraystretch}{1.20}
\caption{UNR Support Levels}
\label{table:UNR_level}
\begin{tabular}{c|c|l|l}
\hline
Level & \begin{tabular}[c]{@{}c@{}}PUT Custom Bits\\ Length at Remote\end{tabular} & \multicolumn{1}{c|}{Implementation Specifications} & \multicolumn{1}{c}{Suggestion for Users} \\ \hline
0 & 0 & \begin{tabular}[c]{@{}l@{}}Additional order preserving message is needed to\\ transfer $p$ and $a$.\end{tabular} & \begin{tabular}[c]{@{}l@{}}For correctness verification only,\\ no guarantee of performance.\end{tabular} \\ \hline
1 & 8, 16 & All the bits are used for $p$, and assume $a=-1$. & \begin{tabular}[c]{@{}l@{}}The maximum number of signals is limited.\\ Performance may degrade if the limit is exceeded.\\ Multi-channel is not supported.\end{tabular} \\ \hline
2 & 32 & \begin{tabular}[c]{@{}l@{}}Mode1: All the bits are used for $p$, and $a=-1$.\\ Mode2: User-specified $x$ bits for $p$ and $32-x$ bits for $a$.\end{tabular} & \begin{tabular}[c]{@{}l@{}}Mode1: Multi-channel is not support.\\ Mode2: Multi-channel can be enabled with \\ \hspace{10mm}a limited number of signals and events.\end{tabular} \\ \hline
3 & 64, 128 & Both $p$ and $a$ use half of the bits. & \begin{tabular}[c]{@{}l@{}}Multi-channel Multi-message Aggregated Signal is \\ completely supported in this level.\end{tabular} \\ \hline
4 & 128 & \begin{tabular}[c]{@{}l@{}}64 bits for $p$, 64 bits for $a$. No polling thread required.\\ (Hardware support atomic add after PUT and GET.)\end{tabular} & \begin{tabular}[c]{@{}l@{}}No need to worry about performance degradation caused \\ by polling threads.\end{tabular} \\ \hline
\end{tabular}
\begin{tablenotes}
\item[!] $p$ is a pointer to the signal's counter, and $a$ is an addend that needs to be added to the counter.
If there are $x$ bits for $p$, the maximum number of available signals is $2^x$.
If $x<64$, these bits only store the index of signal, and the polling thread will look up the signal and execute the add.
\end{tablenotes}
\end{threeparttable}
\end{table*}

\begin{table*}[t]
\centering
\renewcommand{\arraystretch}{1.5}
\caption{The UNR Support Level of High-Performance NICs}
\label{table:NIC_Level}
\begin{tabular}{cccccccc}
\multicolumn{1}{c|}{\multirow{2}{*}{Interface}} & \multicolumn{1}{c|}{\multirow{2}{*}{HPC Interconnect}} & \multicolumn{1}{c|}{\multirow{2}{*}{\begin{tabular}[c]{@{}c@{}}Representative HPC Systems\\ (Highest Ranking on Top500)\end{tabular}}} & \multicolumn{2}{c|}{PUT Custom Bits} & \multicolumn{2}{c|}{GET Custom Bits} & \multirow{2}{*}{UNR Support Level} \\
\multicolumn{1}{c|}{} & \multicolumn{1}{c|}{} & \multicolumn{1}{c|}{} & \multicolumn{1}{c|}{Local} & \multicolumn{1}{c|}{Remote} & \multicolumn{1}{c|}{Local} & \multicolumn{1}{c|}{Remote} &  \\ \hline
Glex & TH Express network\cite{liaoHighPerformanceInterconnect2015} & Tianhe-2A(1), Tianhe-Xingyi & 128 & 128 & 128 & 128 & Level-3 \\ \hline
Verbs\cite{RdmacoreLibibverbsMaster} & Slingshot\cite{desensiInDepthAnalysisSlingshot2020}, Infiniband, RoCE & Frontier(1)\cite{schneiderExascaleEraUs2022}, Summit(1)\cite{stunkelHighspeedNetworksSummit2020} & 64 & 32 & 64 & 0 & Level-2 \\ \hline
uTofu\cite{fujitsuDevelopmentStudioUTofu2021} & Tofu Interconnect & Fugaku(1)\cite{satoCoDesignSystemSupercomputer2022}, K(1)\cite{ajimaTofuInterconnect2018} & 64 & 8 & 64 & 8 & Level-1 \\ \hline
uGNI\cite{enterpriseGNIAPIReference} & Aries Interconnect\cite{faanesCrayCascadeScalable2012} & Piz Daint(3)\cite{swissnationalsupercomputingcentrecscsFactsheetPizDaint2018}, Trinity(6)\cite{losalamosnationallaboratorylanllanlTrinityAdvancedTechnology2020} & 32 & 32 & 32 & 32 & Level-2 \\ \hline
PAMI\cite{kumarPAMIParallelActive2012} & Blue Gene/Q Interconnection\cite{chenIBMBlueGene2011} & Sequoia(1), Mira(3)\cite{baileyBlueGeneSequoia2013} & \multicolumn{2}{c}{64(Shared)} & 64 & 0 & Level-2 \\ \hline
Portals\cite{hudsonPortalsMessagePassing2006} & SeaStar Interconnect\cite{brightwellPortalsSandiaCray} & Kraken(3), Jaguar(6)\cite{yuPerformanceCharacterizationOptimization2008} & Hash & 64 & Hash & 0 & Level-3 \\ \hline
\end{tabular}
\end{table*}

We use a data structure \texttt{signal\_t} to implement MMAS. This data structure contains two signed 64-bit integers, \texttt{num\_event} and \texttt{counter}.
The \texttt{num\_event} represents the number of events needed to trigger the signal. 
The event could be a completion of PUT or GET at a local or remote process.
We set the position of the highest non-zero bit in the maximum \texttt{num\_event} as \texttt{N}.
The \texttt{counter} is divided into three parts. The lower \texttt{N} bits count the remaining event, which will be set to \texttt{num\_event} before use, and the higher \texttt{64-N-1} bits count the remaining sub-messages if a message is divided into multiple sub-messages for transmitting through multiple channels.
$N$ can be set by users according to their needs when the length of custom bits is short.
The one bit in the middle is an event overflow detection bit that indicates if more than \texttt{num\_event} events have been received. It will be discussed in \Cref{sec:bug_avoid}.

We utilize the custom bits within Notifiable RMA Primitives to store a pointer \texttt{p} pointed to the \texttt{counter} and an addend \texttt{a}.
The addend \texttt{a} is added to the \texttt{counter} that \texttt{p} pointed to after the operation is complete.
If a message is transferred through one channel, \texttt{a=-1}.
If a message is fragmented into $K$ sub-messages and transferred through multiple channels,
\texttt{a=-1+[(K-1)<<(N+1)]} in one of the sub-messages, and \texttt{a=(-1)<<(N+1)} for all other sub-messages (left shifting operation pads zeros).
The \texttt{counter} equals zero if and only if all sub-messages have arrived.
Thus, UNR triggers the signal when \texttt{counter} equals zero.
\Cref{fig:MNMMAS} shows an example.

The division can also be used for a single channel to prevent head-of-line blocking in the network\cite{xieExtremescaleDirectNumerical2024} and utilizing both CPU and NIC to implement the intra-node communication.

\subsection{UNR Support Level} \label{sec:UNR_level}

The length of custom bits determines the maximum number of signals, the maximum number of events in a signal, and the maximum number of sub-messages.
We use the length of custom bits in PUT at remote to classify UNR support levels.
The descriptions of each UNR support level are shown in \Cref{table:UNR_level}.
The implementation specifications for each level are defined.
We also provide suggestions for users when using different UNR support level interconnects.
The implementation specifications for GET and local signals are the same.

We use the length of custom bits in PUT at remote rather than GET or local custom bits because
the most important usage of UNR is optimizing two-sided communication with PUT in existing HPC applications.
We recommend using PUT instead of GET since the PUT latency is a one-way delay while the GET latency is round-trip time.
Furthermore, the length of custom bits in PUT is not smaller than that in GET at remote in Notifiable RMA Primitives, as demonstrated in \Cref{table:NIC_Level}.
The length of custom bits at local is usually not smaller than the length of custom bits at remote.
The exception is Portals3.3\cite{hudsonPortalsMessagePassing2006}, which does not provide custom bits at local but does offer memory region and offset for the operation, which can be used as a hash to store $p$ and $a$ at local.

According to \Cref{table:NIC_Level}, the TH Express network offers the best support for UNR, since it provides 128 custom bits for both PUT and GET at both remote and local.
TH Express network can support up to $2^{64}-1$ signals, $2^{31}-1$ sub-messages, and $2^{32}-1$ events in a signal.

In UNR support level 0-3, we utilize a polling thread to retrieve events from NICs and add the addends $a$ to the \texttt{counter} that $p$ points to.
We will show the polling thread may affect the performance of an application in \Cref{sec:powerllel_evaluation}.
To solve this problem, we define level-4, which requires 128 custom bits (in both PUT and GET at both local and remote), and offloading the atomic add \texttt{*p+=a} to hardware.
Since the next-generation TH Express NIC has supported RMA and atomic operations, and level-4 needs a combination of these two operations, we are suggesting the vendor add this feature \controlline{0}{}{to the next-generation interconnect }to achieve hardware-software co-design.
Using system-reserved cores on Fugaku\cite{satoCoDesignSystemSupercomputer2022} or using CPU cores in Bluefield SmartNIC\cite{sureshNovelFrameworkEfficient2023} can also achieve such offload, but both of them have limited custom bits.

\subsection{Bug-avoiding Interface} \label{sec:bug_avoid}


The synchronization error and remote address offset calculation error are two common issues when using RMA operations. The former occurs due to incorrect pre-synchronization and leads to overwriting useful data or getting incorrect data.
The signal in UNR can be used for two purposes: (1) indicating the completion of RMA operations, and (2) detecting synchronization errors in the application.
Users have to call \texttt{UNR\_Sig\_Reset(sig)} to set \texttt{counter} as \texttt{num\_event} each time before using the signal. 
Before the setting, UNR will check if \texttt{counter} equals zero. If \texttt{counter} is not zero, one or more messages have arrived earlier than expected, and UNR will warn the user.
Thus, \texttt{UNR\_Sig\_Reset(sig)} should be called after all corresponding memory regions are ready for RMA operations to check for application synchronization errors.
Although there's a slight chance that errors might go undetected if an RMA operation starts before \texttt{UNR\_Sig\_Reset(sig)} is called and finishes after the reset, this check can still prevent errors in most instances. 
Rerun the application multiple times can also help trigger the warning in UNR and detect the problem for such edge cases.
Besides, \texttt{UNR\_Sig\_Wait(sig)} will check the event overflow detect bit introduced in \Cref{sec:MMAS}.
The bit will become \texttt{1} if more than \texttt{num\_event} events have been received due to the rules of two's complement.

\begin{algorithm}[t]
\caption{Two-sided Communication to be Optimized}
\label{algorithm:before_blk}
\begin{algorithmic}[1]
    \ForAll{iters} 
    \State ...
        \If{$rank=sender$}
            \State MPI\_Send($send\_buf+\textcolor{red}{\mathbf{f(x)}}$,\,$size$)
        \ElsIf{$rank=receiver$}
            \State MPI\_Recv($recv\_buf+\textcolor{red}{\mathbf{g(y)}}$,\,$size$)
        \EndIf
    \State ...
    \EndFor
\end{algorithmic}
\end{algorithm}

To address the issue of remote address offset calculation errors,
we design a data handle BLK in the UNR to avoid calculating remote address offset.
When using UNR, users have to register a memory region before using it in RMA operations.
A BLK represents a block of data in a registered memory region.
BLK stores information for remote processes to access the data block, including the rank of the local process, the handle of registered memory, the offset of the block corresponding to registered memory, and the size of data block.
BLK allows users to use UNR in a fashion similar to two-sided communication operations with a slightly extra overhead to transmit BLK before entering the main loop.
Note that UNR does not register memory for each block because the number of registered memory regions is limited in some systems.
We suggest users register memory as large as possible and then divide it into BLKs.

Code \ref{algorithm:before_blk} and \ref{algorithm:after_blk} present an example of optimizing two-sided communication using UNR.
Code \ref{algorithm:before_blk} uses a common MPI two-sided communication. We assume the addresses of the sending and receiving buffer are complex and represented using $\mathbf{f(x)}$ and $\mathbf{g(y)}$.
Code \ref{algorithm:after_blk} shows how to use UNR PUT to optimize it.
Line 2 and 8 show the registration of memory region.
Line 3 and 9 initialize two signals which will be triggered after 1 event.
Line 4 and 10 declare parts of the memory regions as sending and receiving blocks respectively, and bind the signals to the blocks, which will be triggered after the completion of message sending and receiving, respectively.
Line 6 and 12 transfer the receiving block from the receiver to the sender.
In line 17, UNR extracts the information of remote address from remote block and performs the PUT.
The Receiver waits for the signal to be triggered at line 21.
After the buffer can be reused, the receiver resets the receive signal in line 23.
The sender waits for the completion of sending in line 18, and only after that can the buffer be modified.
The sender also has to reset the signal before its next use, at line 19.

Although the code is longer after optimization, the bug-avoiding interfaces prevent \controlline{1}{}{some common }bugs in programming, which may cost endless time to debug, according to our experience.
UNR also provides other flexible interfaces for users.
For example, \texttt{UNR\_RMA\_Plan()} can record a series of PUT and GET before entering the main loop of the application.
The recorded RMA operations will be executed each time calling \texttt{UNR\_Plan\_Start()}.
The signal can also be specified when planning or calling \texttt{UNR\_Put()} or \texttt{UNR\_Get()}, instead of binding the signal to BLKs like line 4 and 10 in Code \ref{algorithm:after_blk}.
UNR additionally provides interfaces that users need to use to calculate the address offset, like normal one-sided communication.

\begin{algorithm}[t]
\caption{Optimized Code 1 with UNR}
\label{algorithm:after_blk}
\begin{algorithmic}[1]
    \If{$rank=sender$}
        \State $mr \gets$ UNR\_Mem\_Reg($send\_buf$,\,$buf\_size$)
        \State $send\_sig \gets$ UNR\_Sig\_Init(1) \Comment{Trigger after 1 event}
        \State $send\_blk \gets$ UNR\_Blk\_Init($mr$,$\textcolor{red}{\mathbf{f(x)}}$,$size$,$send\_sig$)
        \State \Comment{Trigger the signal if $send\_blk$ finishes sending}
        \State MPI\_Recv($rmt\_blk$) \Comment{Get remote receiving address}
    \ElsIf{$rank=receiver$}
        \State $mr \gets$ UNR\_Mem\_Reg($recv\_buf$,\,$buf\_size$)
        \State $recv\_sig \gets$ UNR\_Sig\_Init(1) \Comment{Trigger after 1 event}
        \State $recv\_blk \gets$ UNR\_Blk\_Init($mr$,$\textcolor{red}{\mathbf{g(y)}}$,$size$,$recv\_sig$)
        \State \Comment{Trigger the signal if $recv\_blk$ finishes receiving}
        \State MPI\_Send($recv\_blk$) \Comment{Send receiving address}
    \EndIf

    \ForAll{iters} 
    \State ... \Comment{Perform pre-synchronization}
        \If{$rank=sender$}
            \State UNR\_Put($send\_blk$, $rmt\_blk$)
            \State UNR\_Sig\_Wait($send\_sig$)
            \State UNR\_Sig\_Reset($send\_sig$)
        \ElsIf{$rank=receiver$}
            \State UNR\_Sig\_Wait($recv\_sig$)
            \State ...  \Comment{Use the received data}
            \State UNR\_Sig\_Reset($recv\_sig$) \Comment{After buffer is ready}
        \EndIf
    \State ...
    \EndFor
\end{algorithmic}
\end{algorithm}

\begin{figure*}[t]
  \centering
  \includegraphics[width=1.0\linewidth]{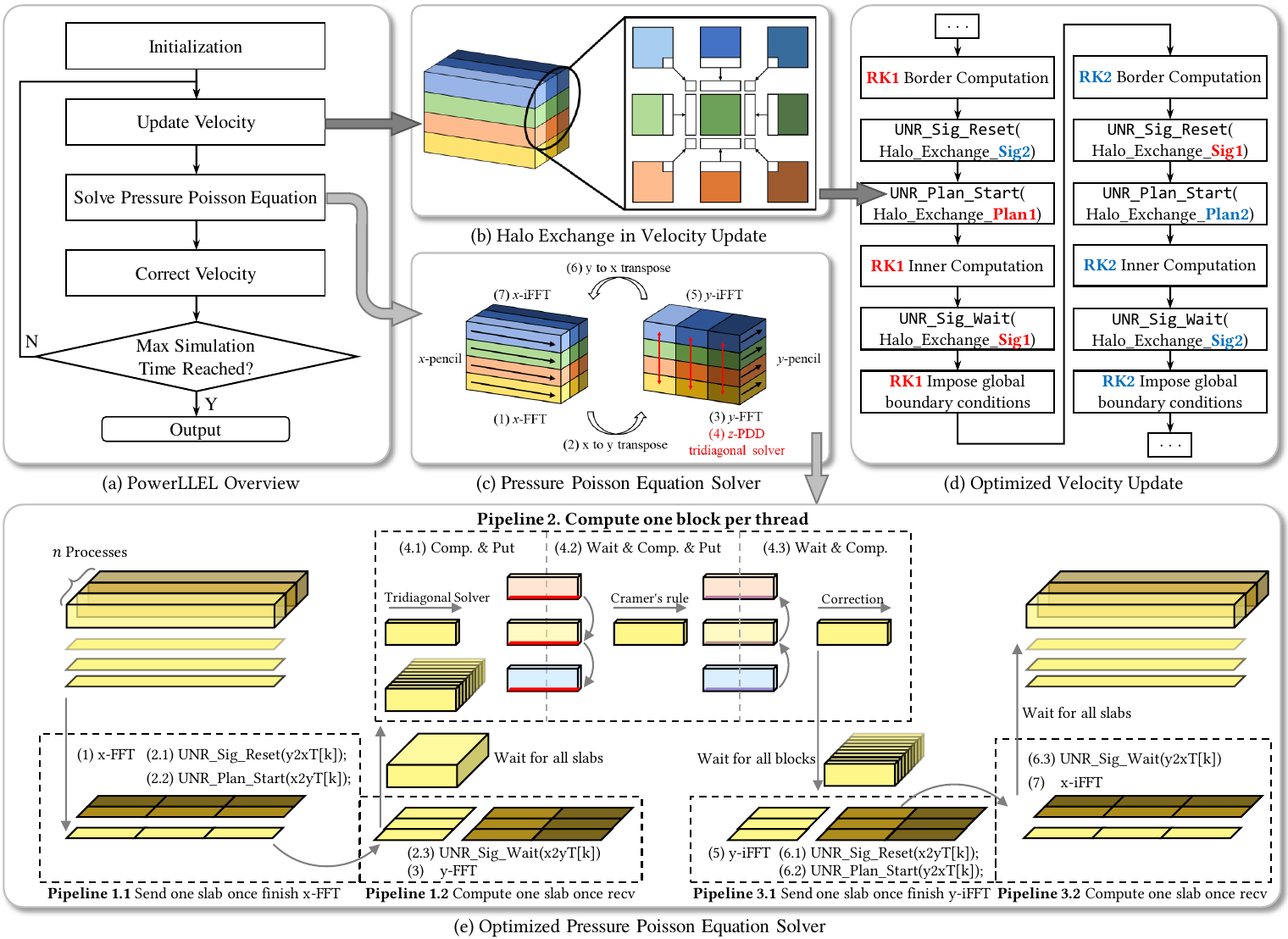}
  \caption{Optimizing PowerLLEL using UNR}
  \label{fig:powerllel}
\end{figure*}

\subsection{Further Analysis and Discussions}

\subsubsection{Reasons for not providing custom bits directly}
On the one hand, it’s cumbersome to let users handle custom bits of varying lengths.
On the other hand, the application may lose its portability when migrating to a new platform with shorter custom bits.

\subsubsection{Optimizations for intra-node communication}
Currently, UNR implements a channel that uses MPI for intra-node communication. 
This allows UNR to use the vendor-optimized intra-node communication method, which is usually configured in the MPI library.
For extreme intra-node communication performance, UNR Transport Channel based on KNEM\cite{goglinKNEMGenericScalable2013} or XPMEM\cite{woodacreSGIAltixTM3000} can be implemented.
UNR can utilize CPU and NICs for intra-code communication\controlline{1}{}{ due to MMAS}, depending on the channel configuration, which is similar to the design of UCX\cite{shamisUCXOpenSource2015}.

\subsubsection{Collective operation interfaces} \label{sec:collective}
UNR does not provide collective operation interfaces because its goal is to unify the different Notifiable RMA Primitives on different interconnects.
We suggest two approaches to utilize UNR for collective operations.  
Firstly, communication optimization engineers can define collective communication interfaces and implement the collective algorithms according to the needs of applications case by case.
Alternatively, collective communication can be implemented as acceleration libraries based on UNR.
Some pioneers have implemented collective communication algorithms based on RMA operations in previous work\cite{iakymchukEfficientEventuallyConsistent2021, ahadEfficientAlgorithmsCollective2018}.

\subsubsection{Limitation}
Firstly, UNR cannot be used with programming models in which an adaptive runtime system is free to migrate its computing task to any processing element, such as Charm++\cite{acunParallelProgrammingMigratable2014}. 
Secondly, when transmitting small messages, users have to pack and unpack them to avoid performance decrease caused by throughput limitation.
Finally, atomic operations are not added to the UNR due to the varying support for atomic operations across different platforms.

\section{UNR Best Practice} \label{sec:best_practice}

In this section, we discuss our experience in optimizing a computational fluid dynamics (CFD) application, PowerLLEL, using UNR and the best practices we learned in this procedure.
Firstly, in \Cref{sec:sync_free}, we introduce the most critical step in optimization using UNR: synchronization-free condition construction. 
We then introduce the computational kernels and communication patterns of the PowerLLEL application in \Cref{sec:PowerLLEL}.
Finally, we present our optimizations for PowerLLEL using UNR in \Cref{sec:PowerLLEL_opt}.

\subsection{Removing Synchronizations} \label{sec:sync_free}
To construct a synchronization-free condition, both the pre-synchronization and post-synchronization need to be removed. 
The notifiability of UNR frees the users from post-synchronization. 
However, users still need to use pre-synchronization or other mechanisms with the same effect to ensure that RMA operations are safe to start remotely. 
In this subsection, we will discuss how to remove pre-synchronization.

Precisely speaking, pre-synchronization cannot be removed, but it can be performed implicitly in a previous communication from the target process to the source process.
This implies that any transmission from the target to the source can represent the pre-synchronization in \Cref{fig:NPUT} and \Cref{fig:NGET}.
Therefore, we aim to identify such transmissions from target to source before RMA operations and ensure the corresponding buffer for RMA operations is ready before the transmission.
Collective operations like Barrier and All-to-All can also serve as pre-synchronization since both the source and the target processes are involved and implicitly synchronized.
The \texttt{UNR\_SIGNAL\_RESET(sig)} introduced in \Cref{sec:bug_avoid} can help users to check the synchronization errors during removing pre-synchronization.
If pre-synchronization cannot be performed implicitly in a previous communication, users have to construct such transmission from the target process to the source process manually.

Unlike micro-benchmarks that usually use only a few communication operations, real-world scientific applications contain more communication steps and different communication operations. We argue that it is likely to find opportunities for hiding pre-synchronization in earlier communication operations in real-world applications. 
We will show the real cases in the following subsections.

\subsection{Introduction of PowerLLEL} \label{sec:PowerLLEL}

PowerLLEL\cite{xieLowCommunicationOverheadParallelDNS2021} is a high-performance finite difference solver for incompressible turbulent flows.
It is developed and optimized for massively parallel Direct Numerical Simulation (DNS) of the turbulent channel flow at high Reynolds numbers and it has demonstrated good scalability on multiple supercomputers. 
In PowerLLEL, the Navier-Stokes (NS) equations are discretized on a 3D Cartesian grid, using a second/fourth-order finite difference scheme.
Time advancement is performed with a second-order Runge-Kutta scheme.
The Pressure Poisson Equation (PPE) is efficiently solved by the FFT-based direct method, combined with a parallel tridiagonal solver based on the Parallel Diagonal Dominant (PDD) algorithm.

Codes for large-scale turbulence DNS are communication-intensive, which is fully reflected in PowerLLEL.
As shown in \Cref{fig:powerllel}a, the computational flow of PowerLLEL mainly consists of three steps: 
updating velocity using the second-order Runge-Kutta scheme, 
solving PPE with FFT and the PDD solver, 
and correcting velocity.
On the one hand, the velocity is updated in a stencil-like manner in every Runge-Kutta substep, which results in halo exchange, a typical communication pattern with spatial locality, as shown in \Cref{fig:powerllel}b.
On the other hand, performing FFT in a commonly used pencil-like 2D domain decomposition requires global transpositions with considerable communication cost, as shown in \Cref{fig:powerllel}c.
These two communication patterns are the scalability bottlenecks of PowerLLEL, so we will focus on their communication optimizations.

\subsection{Optimizing PowerLLEL using UNR}
\label{sec:PowerLLEL_opt}

\begin{algorithm}[t]
\caption{MPI Conversion Interfaces in PowerLLEL}
\label{list:mpi_convert}
\begin{algorithmic}[1]
    \State unr\_plan\_t \textbf{MPI\_Isend\_Convert}(mem\_h, offset, count,\textcolor{white}{dd}
                \textcolor{white}{dd}type, dst, tag, comm, mpi\_request, send\_finish\_sig);
    \State void \textbf{MPI\_Irecv\_Convert}(mem\_h, offset, count,\textcolor{white}{dddddd}
                \textcolor{white}{dd}type, src, tag, comm, mpi\_request, recv\_finish\_sig);
    \State unr\_plan\_t \textbf{MPI\_Alltoallv\_Convert}(\textcolor{white}{ddddddddddddddddddd}
        \textcolor{white}{dd}send\_mem\_h, send\_counts[], send\_displs[], send\_type,
        \textcolor{white}{dd}recv\_mem\_h, recv\_counts[], recv\_displs[], recv\_type,
        \textcolor{white}{dd}comm, send\_finish\_sig, recv\_finish\_sig);
    \State unr\_plan\_t \textbf{MPI\_Sendrecv\_Convert}(\textcolor{white}{dddddddddddddddddddd}
        \textcolor{white}{dd}send\_mem\_h,\,send\_offset,\,send\_count,\,send\_type,\,dst,
        \textcolor{white}{dd}recv\_mem\_h,\,recv\_offset,\,recv\_count,\,recv\_type,\,src,
        \textcolor{white}{dd}comm, send\_finish\_sig, recv\_finish\_sig);
\end{algorithmic}
\end{algorithm}

\begin{table*}[t]
\centering
\renewcommand{\arraystretch}{1.5}
\caption{Experiment Platforms Specifications.}
\label{table:systems}
\begin{tabular}{cccc}
System (abbreviation, deployed year)                            & CPU                                   & NIC(s)                               & Used nodes  \\ 
\hline
Tianhe-Xingyi Supercomputing System (\textbf{TH-XY}, 2024)      & 2$\times$~Multi-core CPU              & 2$\times$200Gbps new TH Express NICs & 1728        \\ 
\hline
Tianhe-2A Supercomputing System(\textbf{TH-2A}, 2013)           & 2$\times$ Xeon~E5-2692 v2 12-core CPU & 114Gbps TH Express NIC               & 192         \\ 
\hline
HPC system interconnected by Infiniband (\textbf{HPC-IB}, 2019) & 2$\times$ Xeon Gold 6150 18-core CPU  & 100Gbps EDR ConnectX-5 NIC           & 24          \\ 
\hline
HPC system interconnected by RoCE (\textbf{HPC-RoCE}, 2019)     & 2$\times$ Xeon Gold 6150 18-core CPU  & 25Gbps ConnectX-4 Lx NIC             & 12          \\
\hline
\end{tabular}
\end{table*}

\begin{figure*}[t]
    \centering
    \includegraphics[width=0.8\linewidth]{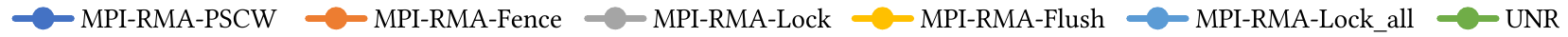}
    \hfill
    \subfloat[TH-XY\label{fig:thxy_lat}]{\includegraphics[width=0.25\linewidth]{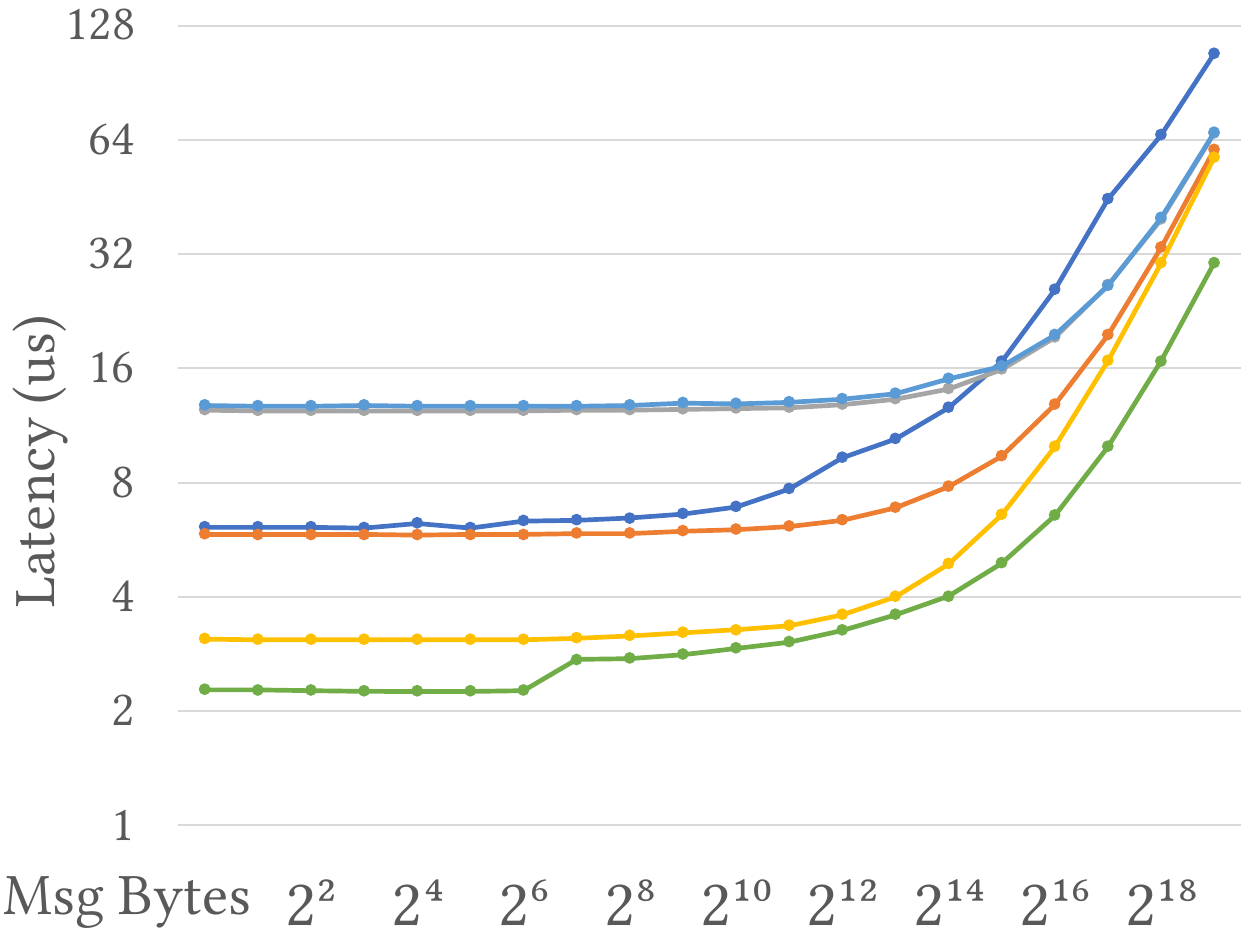}}
    \subfloat[TH-2A\label{fig:th2a_scale}]{\includegraphics[width=0.25\linewidth]{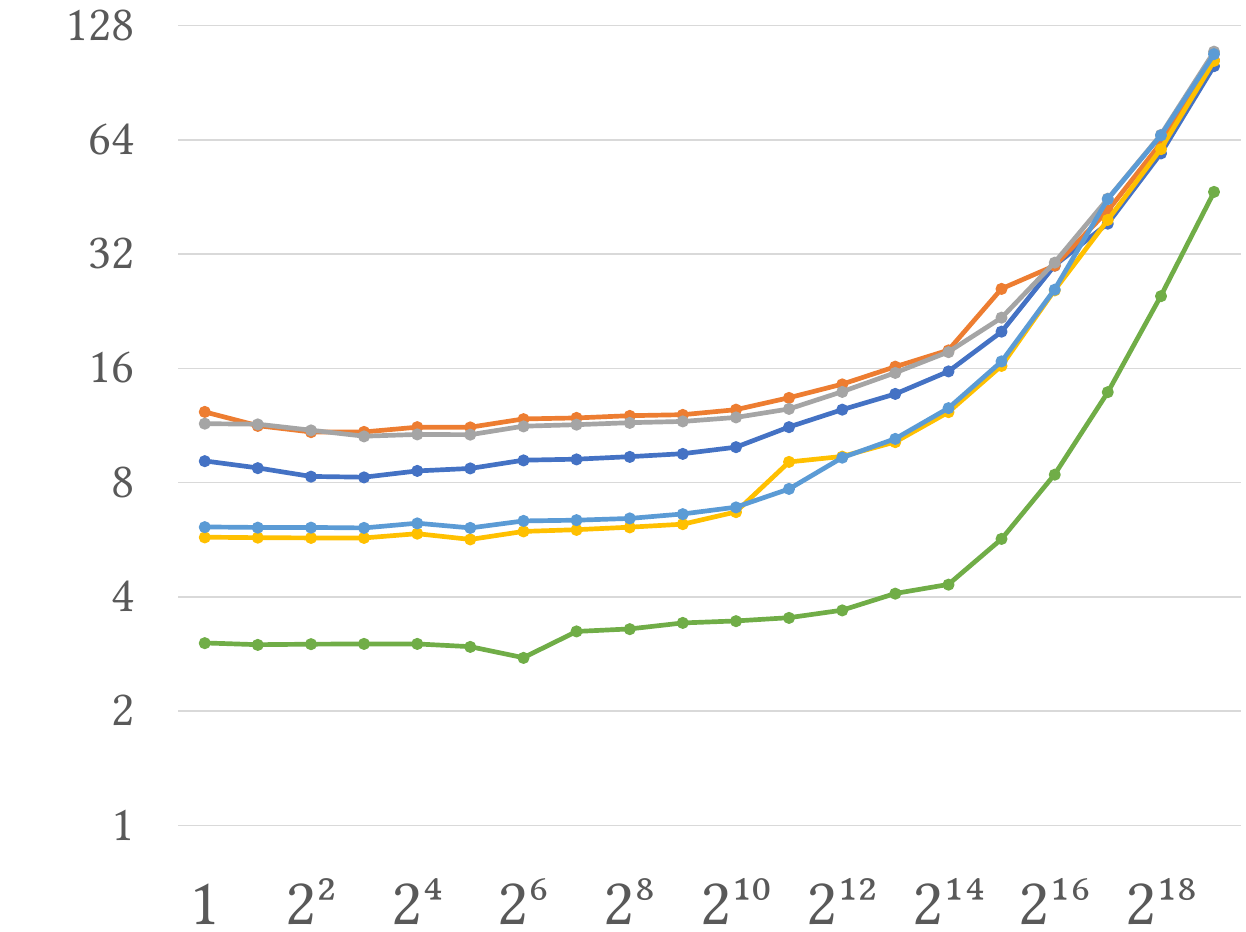}}
    \subfloat[HPC-IB\label{fig:thxy_lat}]{\includegraphics[width=0.25\linewidth]{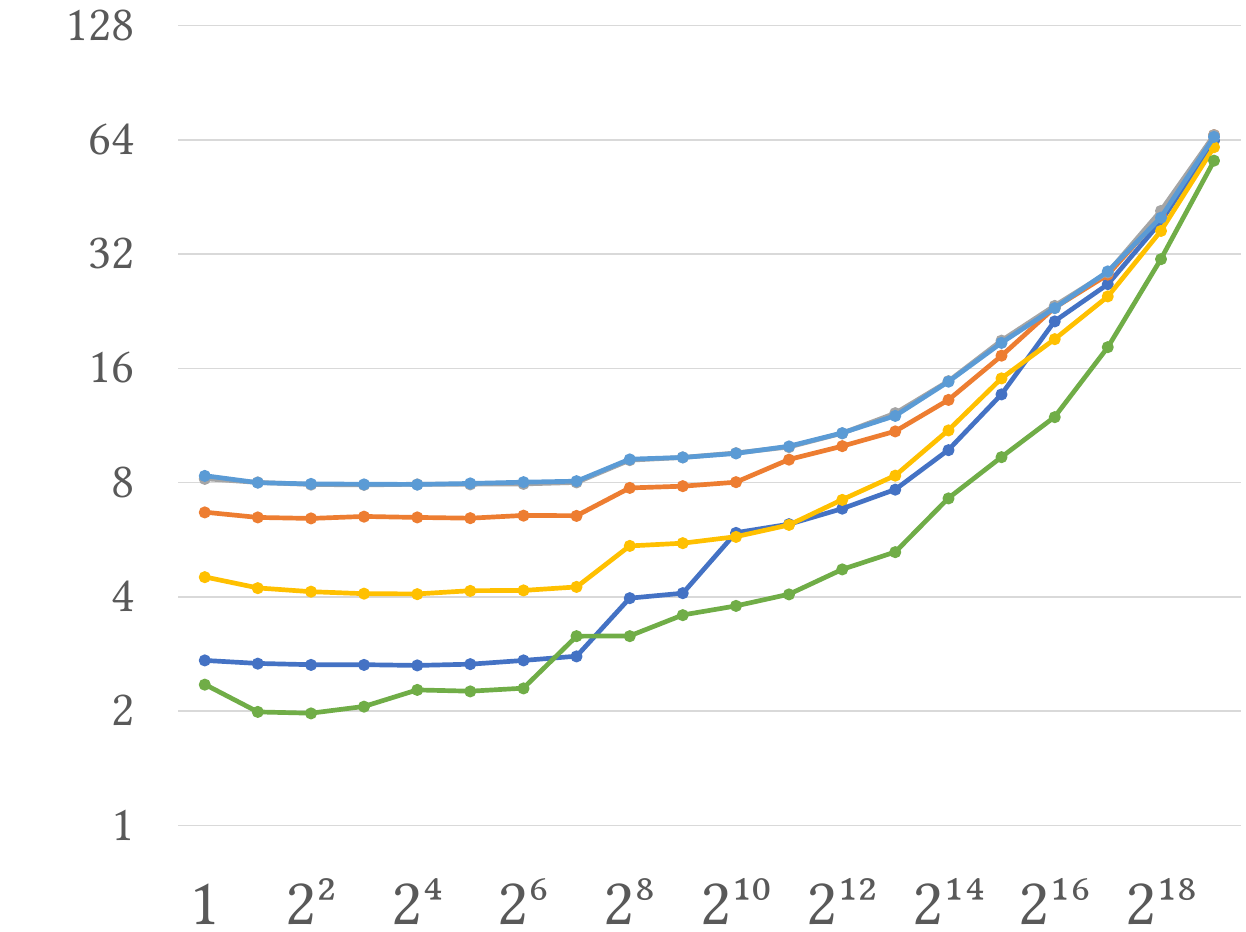}}
    \subfloat[HPC-RoCE\label{fig:th2a_scale}]{\includegraphics[width=0.25\linewidth]{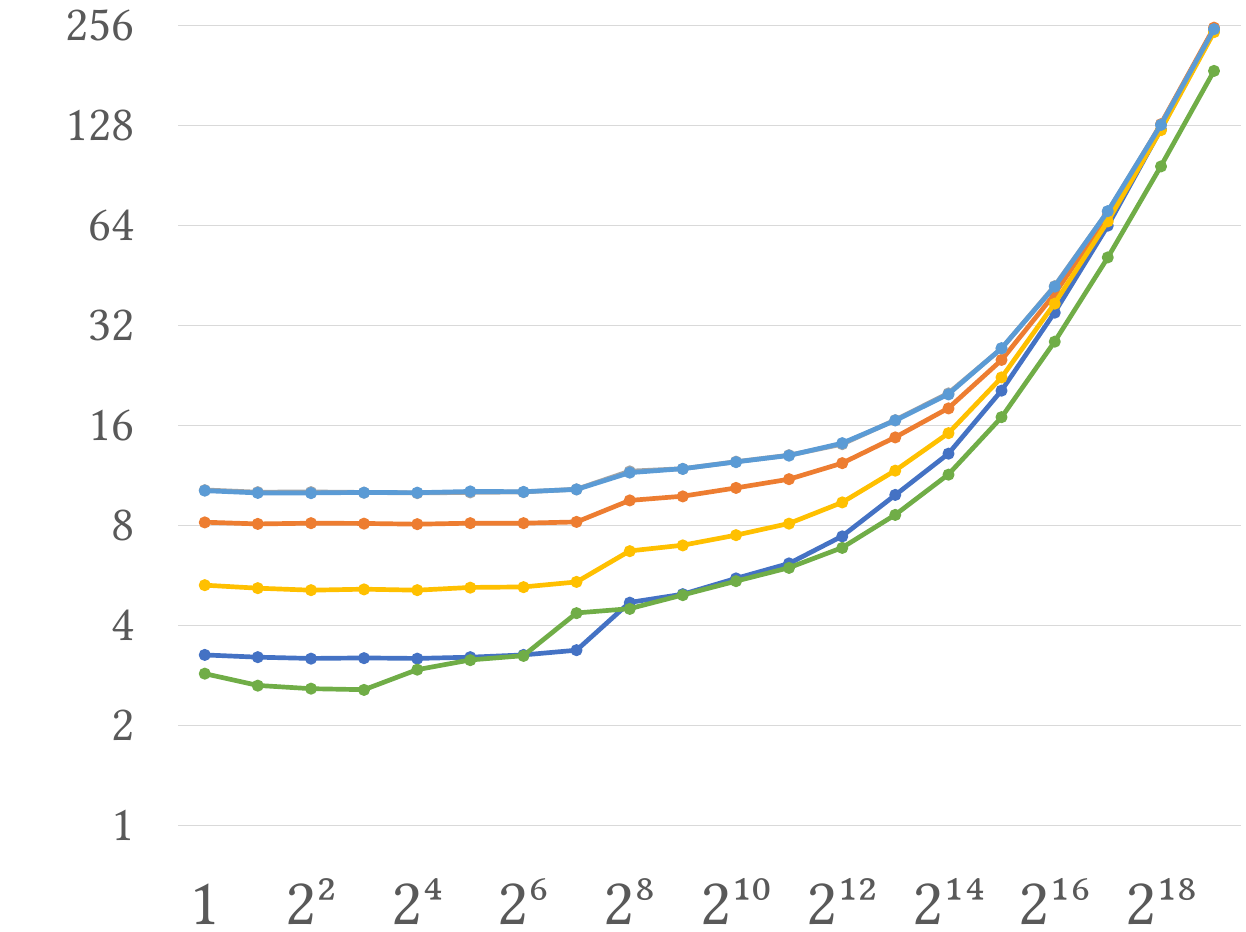}}
    \hfill
    \caption{Latency Test.}
    \label{fig:latency}
\end{figure*}

\subsubsection{Optimization in Velocity Update}
The second-order Runge-Kutta scheme contains two steps, RK1 and RK2, with each step using the same halo exchange operation for different buffer.
Therefore, both steps can be the pre-synchronization of each other, and we can safely remove all explicit pre-synchronization in the velocity update.
\Cref{fig:powerllel}d shows the optimized velocity update.
Before starting the halo exchange of a step, we prepare the buffer and reset the receive signal of another step, which can avoid the synchronization error.

To help the PowerLLEL developer replace two-sided MPI operations with \controlline{0}{}{interfaces of }UNR, we implement \texttt{MPI\_Isend/Irecv\_Convert()} in PowerLLEL with UNR interfaces, as shown in Code \ref{list:mpi_convert} line 1-2.
The parameters of these two functions are similar to \texttt{MPI\_Send/Recv\_Init()}, except that the memory address is replaced by a registered memory handle and local offset.
The completion of \texttt{mpi\_request} represents the exchange of address information has finished, and becomes a transmission task in the \texttt{unr\_plan\_t}.
The \texttt{send/recv\_finish\_sig} are the signals that will be triggered after the transmission\controlline{1}{}{ is complete}.

We design the convert functions for two reasons.
Firstly, the MPI data types defined by PowerLLEL developers represent non-contiguous data structures that contain various physical quantities to be transferred.
Massive conversion from local offset to remote offset usually causes bugs.
These convert functions can extract the entry from defined MPI data types and exchange UNR BLK in the initialization.
Secondly, the halo exchange parameters, including data size, pattern, and the number of halo layers, may be changed in future development.
We try to avoid recalculating the remote addresses manually whenever the parameters are changed.

\subsubsection{Optimization in PPE solver}
We remove all the pre-synchronization in PPE solver and combine it with a pipeline optimization.
The optimized PPE solver is shown in \Cref{fig:powerllel}e, where each dotted box represents a task for one thread. 

The PPE solver contains a transposition from x-pencil to y-pencil and a transposition from y-pencil to x-pencil, each step includes an All-to-All communication with dedicated buffer.
Both transpositions can be the pre-synchronization of each other, as shown in Pipeline 1 and 3.
Each thread sends a slab of data once the corresponding data finishes its computing.
Once a slab of data is received, a thread can consume the data.
Similarly, we implement \texttt{MPI\_Alltoall\_Convert()} in PowerLLEL for developers, as shown in Code \ref{list:mpi_convert} line 3.

The tridiagonal solver in the PPE solver contains a transmission to the bottom neighbor and a transmission to the top neighbor, as shown in Pipeline 2 of \Cref{fig:powerllel}e.
Both transmissions can also be the pre-synchronization of each other.
Similarly, the computation and communication are pipelined and overlapped.
We also implement \texttt{MPI\_Sendrecv\_Convert()} in PowerLLEL for developers, as shown in Code \ref{list:mpi_convert} line 4.

\subsubsection{Costs of Optimizing with UNR}
It is not the first attempt to optimize PowerLLEL with RMA operations. In our previous work \cite{xieExtremescaleDirectNumerical2024}, PowerLLEL has been optimized for the new Tianhe supercomputing system by Notifiable RMA Primitives.
However, this approach sacrificed portability and took several months to debug synchronization errors and address offset calculation errors. 
The readability of the optimized code is also significantly reduced, and the developers deprecated the old optimization in PowerLLEL updates.
With UNR, we finished the optimizations and debugging of PowerLLEL in a week while maintaining the readability of the optimized code.
Furthermore, the optimized PowerLLEL is now compatible with a wide range of HPC systems with various interconnects.

\begin{figure*}[t]
    \centering
    \subfloat[
        2 Nodes $ \times $ 2 Procs $ \times $ 2 Ping-pongs. 
        \label{fig:pingpong2}]{\includegraphics[width=0.585\linewidth]{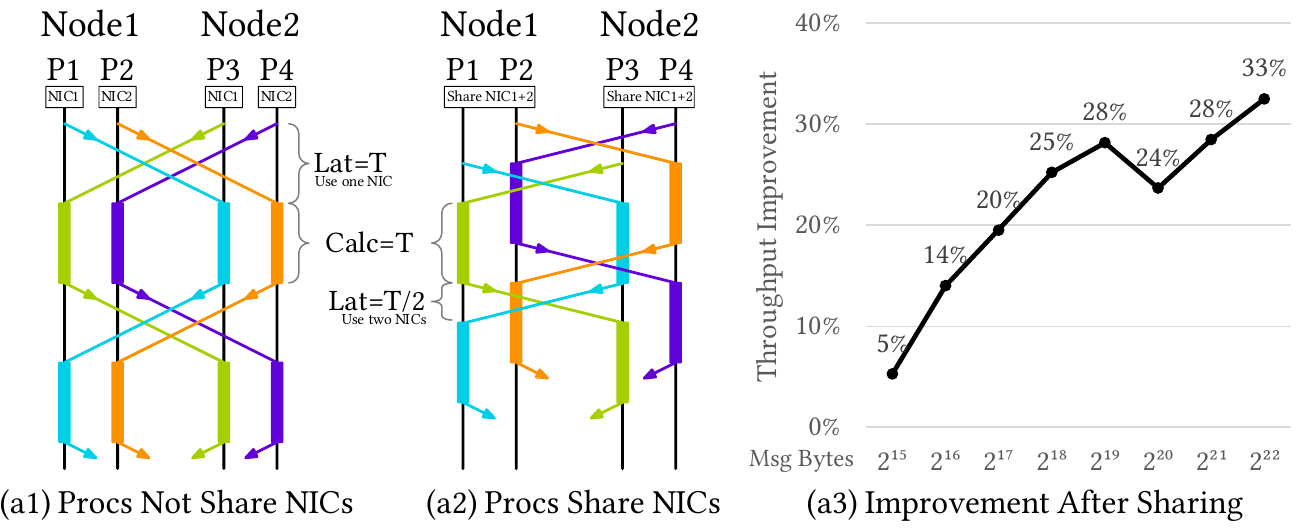}}
    \hfill
    \subfloat[
        2 Nodes $ \times $ 2 Procs $ \times $ 4 Ping-pongs.
        \label{fig:pingpong4}]{\includegraphics[width=0.4\linewidth]{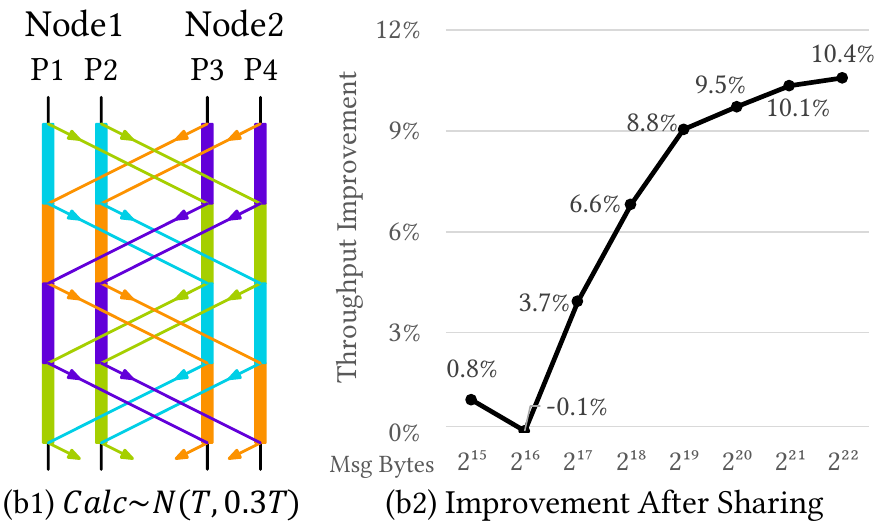}}
    \caption{
        \textbf{UNR Ping-pong Tests with Calculation.}
        Sharing NICs can improve throughput because (a) it allows some messages to be received and calculated in advance, and (b) absorbs the load imbalance in computation.}
    \label{fig:Pingpong}
\end{figure*}

\begin{figure*}[t]
    \centering
    \includegraphics[width=0.992\linewidth]{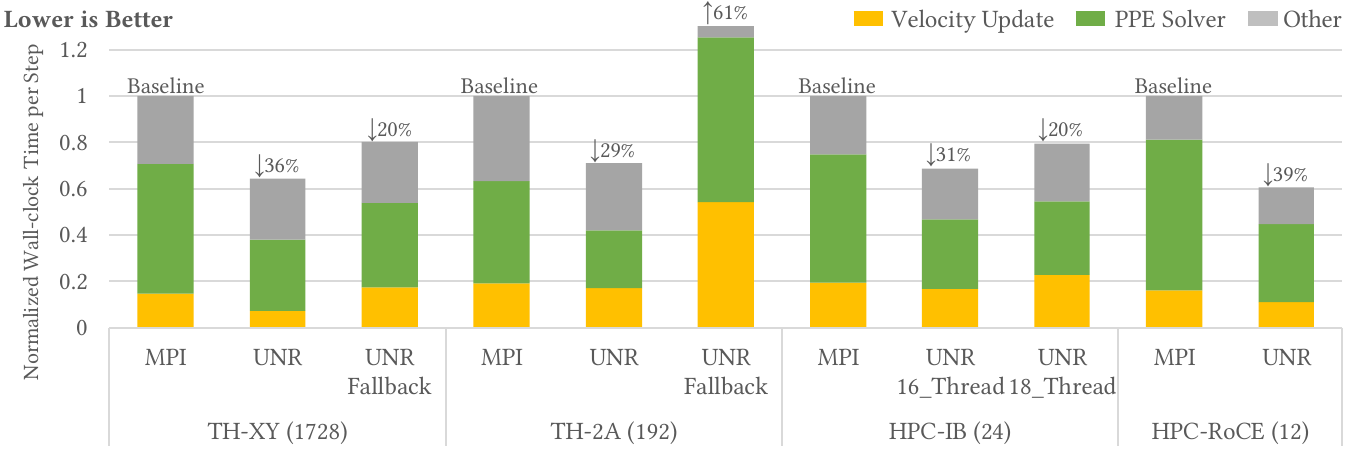}
    \hfill
    \caption{\textbf{PowerLLEL Performance Improvements on Four HPC systems (Number of Nodes Used).} (1) UNR can accelerate PowerLLEL on all four different HPC systems. (2) UNR provides a fallback MPI channel, which can accelerate application in some cases, as shown in TH-XY UNR Fallback, but sometimes not, as shown in TH-2A UNR Fallback. (3) The polling thread does affect PowerLLEL performance. Reserving dedicated CPU cores for polling thread can improve performance, as shown in 16\_Thread and 18\_Thread in HPC-IB.}
    \label{fig:PowerLLEL_test}
\end{figure*}

\section{Evaluation} \label{sec:evaluation}

In this section, we focus on evaluating the portability and performance of UNR.
In \Cref{sec:platforms}, we introduce four HPC systems used for experiments.
In \Cref{sec:benchmark}, we show the basic performance of UNR.
We first compare the latency of UNR with MPI-RMA operations and then evaluate the performance improvements brought by multi-NIC aggregation.
In \Cref{sec:powerllel_evaluation}, we show the performance improvement of PowerLLEL after optimization.

\subsection{Experiment Platforms} \label{sec:platforms}

We evaluate UNR on four HPC systems. \Cref{table:systems} lists the key features of these systems. 
Each of the systems adopts a different interconnect.
The TH Express on TH-XY is a new generation interconnect system, with two NICs per node and a larger bandwidth compared to the interconnect on TH-2A. The network programming interfaces on TH-XY are also slightly different from those on TH-2A.
Both the HPC-IB and HPC-RoCE systems use the Verbs interface, with some small differences in initialization. 
All aforementioned programming interface differences are handled by UNR; no change is needed for the application code. 


\subsection{Micro-benchmarks} \label{sec:benchmark}

We use two micro-benchmarks to evaluate different properties of UNR. 
The first micro-benchmark compares the latency of UNR with the latency of MPI-RMA operations using different synchronization schemes on two nodes of each system.
A simple ping-pong test evaluates the latency of UNR. The latency of MPI-RMA operations is evaluated by OSU Micro-Benchmarks\cite{MVAPICHBenchmarks}.
We select the MPI implementations suggested and tuned by system administrators\controlline{1}{}{, i.e. MPICH-4.1.2, MPICH-3.2.1, OpenMPI-3.1.4, and OpenMPI-3.1.4, respectively}.
The results are shown in \Cref{fig:latency}.
The latency of UNR is lower than MPI-RMA in most cases.
It is worth noting that the latency of MPI-RMA operation synchronized by PSCW is close to UNR or even better in some cases on HPC-IB and HPC-RoCE.
This is because PSCW (Post-Start-Complete-Wait) can be implemented like two-sided communication, leading to poor computation-communication overlap.
\cite{belliNotifiedAccessExtending2015} and \cite{sergentEfficientNotificationsMPI2019} have shown it may damage applications' performance.

The second micro-benchmark measures the performance improvement of multi-NIC aggregation on TH-XY.
HPC programs usually use multiple processes or threads per node on modern HPC systems. 
Multiple NICs on the same node can be utilized by multiple processes/threads, with each process/thread using only one NIC.
The flow completion time can be further reduced if each process/thread can harness multiple NICs. 
To simulate computation-communication overlap, we use ping-pong tests with calculations between receiving previous data and sending the next data.
There are two processes and two NICs on both nodes, with each process executing multiple ping-pongs with a peer process on another node.

\Cref{fig:Pingpong}(a1-a2) shows each process pair executes two ping-pongs.
The latency of transmitting a message is $T$ through one NIC and $T/2$ (ideally) through two NICs.
The throughput can theoretically increase by up to $1/3$ after two processes share dual NICs, 
since sharing NICs allows some messages to be received and calculated in advance. \Cref{fig:Pingpong}(a3) shows the throughput improvement for different message sizes on two nodes of TH-XY.
The larger the message, the greater the throughput improvement.

\Cref{fig:Pingpong}(b1) shows each process pair executes two ping-pongs.
If the calculation time equals \controlline{1}{message transfer latency}{the latency of transferring a message} using one NIC, the throughput cannot be improved because CPUs and NICs are fully utilized.
However, with an unstable calculation time, sharing NICs can absorb the load imbalance in computation and improve the throughput.
\Cref{fig:Pingpong}(b2) shows the performance improvement when calculation time follows a normal distribution $N(T, 0.3T)$.
The throughput increases by about 10\% when the message size is large.





\subsection{PowerLLEL Evaluation} \label{sec:powerllel_evaluation}

\begin{figure}[t]
    \centering
    \includegraphics[width=1.0\linewidth]{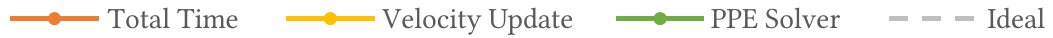}
    \hfill
    \subfloat[TH-2A\label{fig:th2a_scale}]{\includegraphics[width=0.53\linewidth]{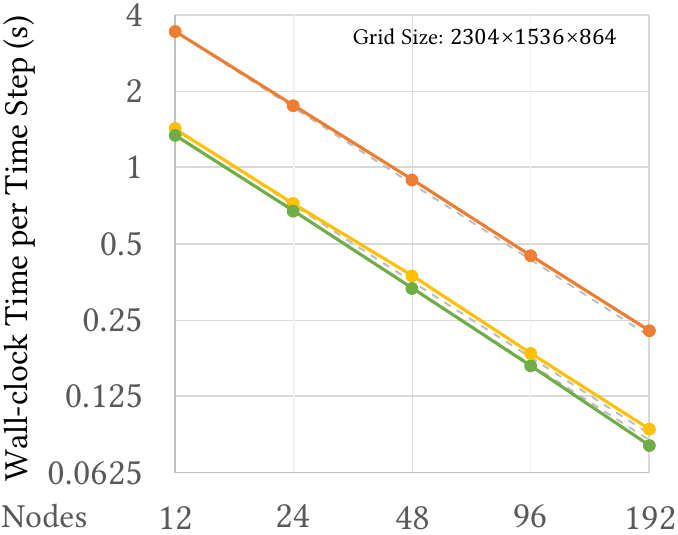}}
    \hfill
    \subfloat[TH-XY\label{fig:thxy_scale}]{\includegraphics[width=0.465\linewidth]{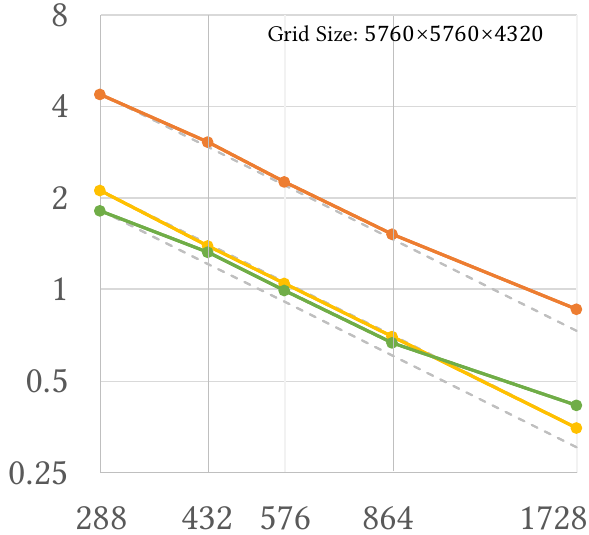}}
    \caption{PowerLLEL Strong Scalability Evaluation.}
    \label{fig:PowerLLEL_scalability}
\end{figure}

\Cref{fig:PowerLLEL_test} shows the PowerLLEL performance improvement on four different HPC systems, tested with grid sizes tailored to fit within the memory constraints of each system.
We use the original MPI version PowerLLEL as the baseline and show runtime breakdowns.
PowerLLEL, powered by the UNR, demonstrates improvements across all four HPC systems, ranging from 29\% to 39\%. UNR has an MPI fallback channel to ensure that UNR-powered applications can still run if no other channel is supported on a system. 
However, the performance of the fallback channel depends on application, interconnect, and MPI implementation.
With UNR fallback channel, PowerLLEL can be accelerated by 20\% on TH-XY, but it slows down by 61\% on TH-2A.

UNR currently has to use a polling thread to poll events from NIC to prevent complete event queue overflow.
However, the polling thread will affect the performance if no dedicated CPU core is reserved for it.
As shown in \Cref{fig:PowerLLEL_test} HPC-IB, if PowerLLEL uses all 18 CPU cores per node with one OpenMP thread per core, a 20\% speedup is achieved. 
If two cores are reserved for the UNR polling thread (due to grid size constraints, we cannot use 17 cores per node), PowerLLEL achieves a better performance: 31\% speedup. 
If no CPU core is reserved for the polling thread, the polling interval needs to be adjusted carefully. A small interval prevents complete event queue overflow and reduces latency, but it competes for more CPU sources and slows down the calculation; and vice versa for a large interval.
The level-4 UNR support discussed in \Cref{sec:UNR_level} is proposed for addressing this issue.

\Cref{fig:PowerLLEL_scalability} shows the strong scalability of PowerLLEL on TH-2A and TH-XY, including the time breakdown of velocity update and PPE solver.
On TH-2A, PowerLLEL with UNR achieves a 95\% parallel efficiency when scaling from 12 nodes to 192 nodes\controlline{1}{}{, with a grid size of $2304\times1536\times864$}.
It achieves an 85\% parallel efficiency when scaling from 288 nodes to 1728 nodes on TH-XY, with a larger grid size of $5760\times5760\times4320$.
From the timing breakdown, the velocity update achieves linear parallel efficiency since the communication time is completely overlapped by computation.
The PPE solver becomes the only performance bottleneck, with only 73\% parallel efficiency.



\section{Conclusion} \label{sec:conclusion}

In this work, we proposed the UNR library to address multi-NIC aggregation, portability, hardware-software co-design, and usability challenges. UNR contains the following core designs:
\begin{itemize}
\item \textbf{Multi-channel Multi-message Aggregated Signal} utilizes the custom bits in Notifiable RMA Primitives as a pointer to the counter and an addend added to the counter, for the purpose of multi-NIC aggregation.
\item \textbf{UNR support levels} defines how well a NIC can support UNR by the length of PUT custom bits at remote, and achieves portability by defining each level's implementation specification. It further requires future NICs to combine RMA and atomic add to replace the polling thread as a hardware-software co-design.
\item \textbf{Bug-avoiding interfaces} provides usability by performing automatical checks on the counter when calling signal-related interfaces and providing users with BLK as a data handle to transmit address information.
\end{itemize}



Our optimizations for the PowerLLEL application present some best practices for using UNR, including performing all pre-synchronization in previous communication steps and avoiding manual calculation of remote address offsets. 

The performance evaluations show promising results of UNR on multiple HPC systems with different interconnects. Micro-benchmarks show two advantages of UNR: (1) achieving lower latency compared to synchronization-based RMA operations in MPI,
and (2) reducing flow completion time and load imbalance by early reception and computation of some messages when using multiple NICs in a node. 
For the PowerLLEL application, UNR delivers 20\% to 39\% speedups on four HPC systems and archives an 85\% parallel efficiency when scaling from 288 to 1728 nodes on the Tianhe-Xingyi supercomputing system.

\section{Future Work} \label{sec:future_work}

We are leveraging compiler techniques to develop an auto-detection pass, and an automatic replacement pass to remove the pre-synchronization.
We plan to use the static rank analysis and dataflow analysis in LLVM IR in the detection pass to determine whether the application meets the conditions for synchronization-free RMA.
For the replacement pass, we will modify the call instructions in LLVM IR to transform original MPI function calls into equivalent UNR interface calls.

We are collaborating with domain-specific developers to adopt UNR in various applications, 
including a brain simulation application with many irregular broadcast operations in each time step for simulating spike broadcasts of neurons. 

With GPU Direct RDMA (GDR) \cite{GPUDirectRDMA}, UNR can be used in a GPU-accelerated system. We are investigating the approaches of supporting other accelerators (for example, the MT-3000\cite{luMT3000HeterogeneousMultizone2022} and AI accelerators) in UNR. 
We hope to accelerate more workloads with UNR in the future, including the training of large language models. 

\section*{Acknowledgment}
Thanks to Yuewen Huang, Hua Huang and Wenkai Shao who gave many valuable comments on this work. And thanks to everyone who has contributed to this work in any way.
This work is supported by the National Key R\&D Program of China under Grant NO. 2022YFB4500304,
the Major Program of Guangdong Basic and Applied Research: 2019B030302002,
National Natural Science Foundation of China (NSFC): 62272499 and 62332021,
Guangdong Province Special Support Program for Cultivating High-Level Talents: 2021TQ06X160.
Yutong Lu is the corresponding author of this paper.


\bibliographystyle{IEEEtran}
\bibliography{IEEEabrv, reference}

\end{document}